# Edge Detection using Stationary Wavelet Transform, HMM, and EM algorithm

S.Anand, K.Nagajothi, K.Nithya
Mepco Schlenk Engineering College, Sivakasi. India

*Abstract*- Abstract- Stationary Wavelet Transform (SWT) is an efficient tool for edge analysis. This paper a new edge detection technique using SWT based Hidden Markov Model (WHMM) along with the expectation-maximization (EM) algorithm is proposed. The SWT coefficients contain a hidden state and they indicate the SWT coefficient fits into an edge model or not. Laplacian and Gaussian model is used to check the information of the state is an edge or no edge. This model is trained by an EM algorithm and the Viterbi algorithm is employed to recover the state. This algorithm can be applied to noisy images efficiently.

I. INTRODUCTION

The edge detection is a very useful preprocessing step in image processing and computer vision-based applications, as it can locate significant variations of gray images [1], [2], [3]. Normally, an edge detection algorithm includes two steps: the image enhancement that estimates image spatial derivatives and the pixel classification that classifies image pixels into two groups, edge or non-edge. Since the edges are key visual features of images for humans, the most straight forward way to detect them is to classify as the edge points whose gray values are larger than a pre-defined threshold. However, it is difficult to precisely pre-define a threshold value for each scale. Nowak described a Bayesian approach to detect edges based on a multiscale hidden Markov model (MHMM) [4], [5], [6]. In his approach, two hidden state values are defined: the state "0" for non-edge and the state "1" for an edge. The multiscale data is an observation of the hidden states. It has been found that the observations of the state "0" are distributed with a low variance Gaussian function since they represent smooth regions and the observations of the state "1" are distributed with a high variance Gaussian function since they represent singular regions. In this, the SWT coefficients in different sub-bands are grouped into a vector. HMM based edge detection have various applications [7], [8], [9], [10].

      In this paper, we propose a new edge detection algorithm based on the undecimated SWT which has the shift-invariant property and efficiently locates the edges. Inspired by the non-decimated wavelet, we propose a novel edge detection technique using the Hidden Markov Chain (HMC) model that can fuse the inter-scale dependence of SWT coefficients.

II. STATISTICAL MODEL FOR INDIVIDUAL SWT COEFFICIENTS:

      The magnitude of a SWT coefficient usually shows high correlations with the nearby ones. This property has been exploited in many SWT-based image processing techniques. However, conventional works consider only the local neighborhood of a coefficient when inferring its hidden state. The transform of a typical signal or image consists of a small number

of large coefficients and a large number of small coefficients. Thus we can roughly model each coefficient as being in one of two states: "high" or "low." If we associate with each state a probability density - say a high-variance, zero-mean density for the "high or big" state and a low-variance, zero- mean density for the "low or small" state – the result is a two-state mixture model for each SWT coefficient.

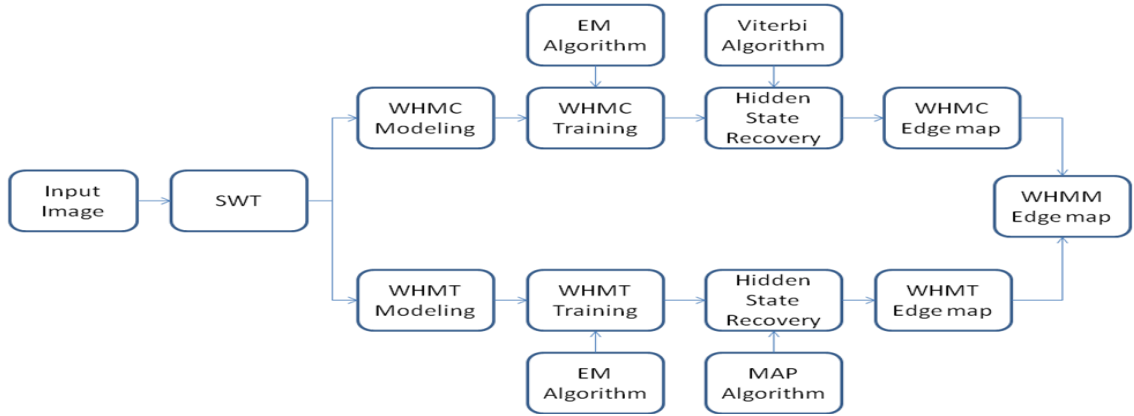

Fig.1. Proposed Methodology

Suppose they have two hidden states: the "small" one $S=0$ and the "big" one $S=1$. There exists a classical zero-mean, two-state Gaussian Mixture Model to represent every coefficient W. Therein, hidden state $S=1$ corresponds to a high-variance Gaussian PDF (Probability Density Function) and $S=0$ corresponds to a low-variance Gaussian PDF. The SWT coefficients with large values correspond to singularity regions in the image since they represent significant signal information. The latter is an indication of the edges. A hidden stochastic state variable s can be defined, which takes two values: either edge (s = 1) or non-edge (s = 0). Then, the corresponding SWT coefficients for s = 0 will satisfy a zero-mean Gaussian distribution with small variance and the corresponding SWT coefficients for s =1 will satisfy a zero-mean Gaussian distribution with a large variance since they represent singularity regions [3]. In the following, all SWT coefficients and their corresponding hidden states for an image are denoted as vector w and s. All SWT coefficients and their corresponding hidden states for a scale j in an image are denoted as vector $w_j$ and $s_j$. For the scale j, we denote a SWT coefficient at a node i as $w_{ij}$ and the corresponding hidden state is denoted as $s_{ij}$. A hidden state at the scale j is described by its probability mass function (pmf)$p(s_{ij})$, which can be denoted by a vector $P_j$

$$P_j = [p(s_{ij}=0) \quad p(s_{ij}=1)]$$

Conditioning on a hidden state $s_{ij}$, the conditional probability density function (pdf) $f(w_{ij} / s_{ij})$ of the SWT coefficients $w_{ij}$ follows a zero-mean Gaussian function. Then, we can have the pdf of the SWT coefficients $w_{ij}$

$$f(w_{ij}) = \sum_{m=0}^{1} f(w_{ij} / s_{ij} = m)p(s_{ij} = m),$$

where m is the state value that is either 0 or 1 and $p(s_{ij} = 0) + p(s_{ij} = 1) = 1$. The formula indicates that $f(w_{ij})$ is a two-state mixture Gaussian function [3]. SWT coefficient exhibits within and across the neighborhood activities which are orientation dependent. However, the model conflicts with the compression property, which dictates the sparse distribution of SWT coefficients. That is to say, only a few coefficients contain most energy and others contribute little to signal energy, as a result, PDF of SWT coefficients should be highly concentrated around zero and heavy-tailed at both sides, all of which cannot be expressed using the previous model. The General Gaussian Distribution (GGD) model can represent peak around zero and long heavy tail approximately

$$f(y) = \frac{p}{2q\Gamma(\frac{1}{p})} \exp(-\left|\frac{y}{q}\right|^p)$$

GGD model has two parameters: standard deviation *q* and shape parameter *p*. Although GGD model can represent peak and heavy tail effectively, its parameter estimation is too complex. So, it proved that for "big" coefficients, the GGD is approximately a Laplacian distribution. As a result, we use an individual Gaussian model for "small" coefficients and Laplacian models for "big" coefficients to simplify the GGD fit. They form a new simple zero-mean, two-state mixture model that can express SWT coefficients effectively.

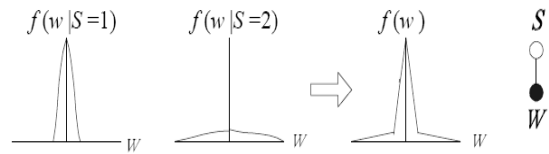

Fig.2: A two-state Gaussian mixture model for a SWT coefficient *W*. We denote the state variable S with a white dot, the SWT coefficient *W* with a closed dot. Illustrated are the Gaussian conditional pdf's for *W* IS as well as the overall mixture pdf for W.

III. THE HMC AND HMT FOR CASUAL DEPENDENCIES:

*Hidden Markov Chain Model:* Connecting the state variables $S_i$ in Fig. 2 horizontally specifies a Markov chain dependency between the state variables within each scale. This model treats SWT state variables as dependent within each scale, but independent from scale to scale.

*HMT Model:* By connecting state variables vertically through scale in Fig. 2, we obtain a graph with tree-structured dependencies between state variables. We call this new model a "tree model" to emphasize the underlying dependencies between parent and child state variables.

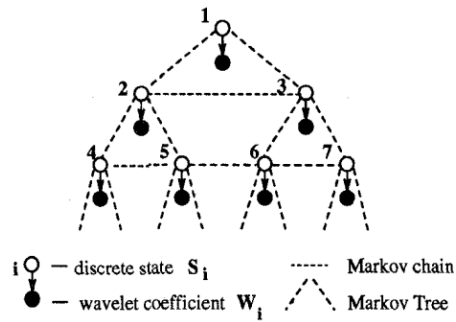

Fig.3: Probabilistic graphs for modeling the statistical dependencies of the coefficients of a SWT transform. Each black node represents a SWT coefficient W. Each white node represents the (hidden) mixture state variable *S,* for W. Removing all dashed connections correspond to the independent Mixture Model. Connecting discrete nodes horizontally across time yields the Hidden Markov Chain Model. Connecting discrete nodes vertically across scale yields the Hidden Markov Tree Model.

## IV. THE WD-HMC AND WD-HMT MODEL TRAINING:

*EXPECTATION-MAXIMIZATION ALGORITHM:*

We adopt an EM algorithm [7] to train both the model. Starting from an initial set of parameter estimates, the EM algorithm iterates between the following two steps:

*WHMC Training*
*Expectation (E step):* Estimate the likelihood (or probability) of the data given the current parameters.
*Maximization (M step):* Select new parameters to maximize the estimated likelihood of the data.

*WHMT Training*
*Upstep:* calculates the conditional likelihoods by transmitting information from states of the fine-scale SWT coefficients to the states of the coarse-scale SWT coefficients [9].
*Down step:* calculates the joint probability by propagating information about the states from the coarse-scale SWT coefficients down to of the fine-scale SWT coefficients [9].

*VITERBI ALGORITHM:*

The Viterbi algorithm [7] chooses the best state sequence that maximizes the likelihood of the state sequence for the given observation sequence based on Maximum-a-Posteriori(MAP). These are the steps,
*Initialization:*

$$\delta_1(i) = p_i\, b_i(o(1))$$
$$\psi_1(i) = 0,\ i = 1, \ldots, N$$

According to the above definition, $\beta_T(i)$ does not exist. This is a formal extension of the below recursion.
*Recursion:*

$$\delta_t(j) = \max_i [\delta_{t-1}(i)\, a_{ij}]\, b_j(o(t))$$
$$\psi_t(j) = \arg\max_i [\delta_{t-1}(i)\, a_{ij}]$$

*Termination:*

$$p^* = \max_i [\delta_T(i)]$$
$$q^*_T = \arg\max_i [\delta_T(i)]$$

*Path (state sequence) backtracking:*
$q^*_t = \psi_{t+1}(q^*_{t+1})$, $t = T - 1,\ T - 2, \ldots, 1$

## V. EXPERIMENTAL RESULTS

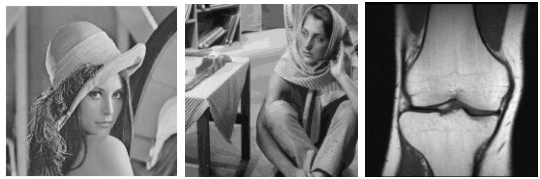

Fig.4. Test images Lena, Barbara, Knee

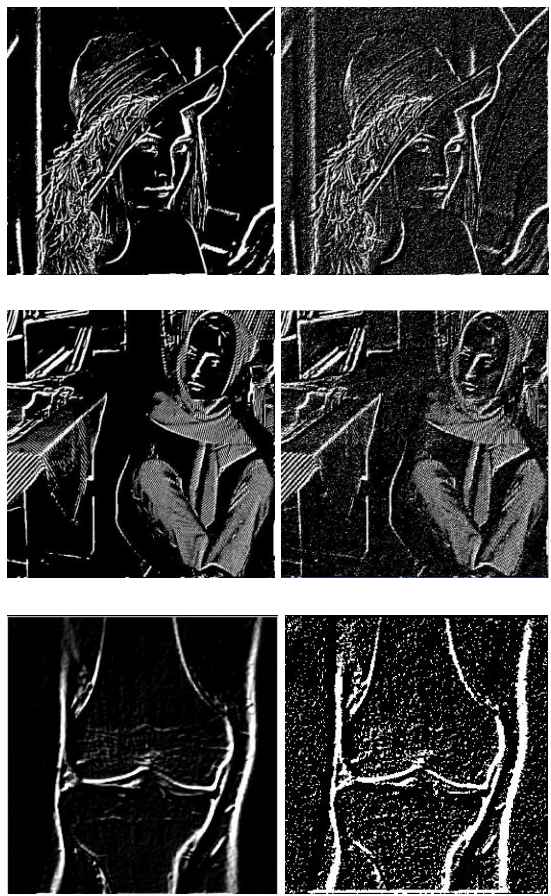

Fig .5 Edge detection results for clean and random noisy images

Table No. 5.1: Performance Comparison

| Methodology | Number of detected edges |
|:---:|:---:|
| Canny | 3972 |
| Log | 4831 |
| Sobel | 879 |
| Roberts | 878 |
| Zero crossing | 4811 |
| WHMM | 5096 |

VI. CONCLUSION

The multi-resolution property of the SWT transform has led to its efficiency in singularity detection as a multi-scale tool. A new edge detection algorithm WD-HMM using HMM model based on the shift invariant SWT transform is proposed. Experimental results show that the proposed method is an efficient and accurate edge detecting tool for clean and noisy images.

VII. REFERENCES

[1]. Anand, S., T. Thivya, and S. Jeeva. "Edge detection using directional filter bank." International Journal of Applied Information Systems 1 (2012): 21-27.
[2]. Anand, S., and NM Mary Sindhuja. "Spot edge detection in microarray images using balanced GHM multiwavelet." *2009 International Conference on Control, Automation, Communication and Energy Conservation*. IEEE, 2009.
[3]. Anand, S., and H. Abirami. "Non-separable steerable filters for edge detection." *2011 International Conference on Signal Processing, Communication, Computing and Networking Technologies*. IEEE, 2011.
[4]. Zhang, Renqi, Wanli Ouyang, and Wai-Kuen Cham. "Image edge detection using hidden Markov chain model based on the non-decimated wavelet." *2008 Second International Conference on Future Generation Communication and Networking Symposia*. Vol. 3. IEEE, 2008.
[5]. Sun, Junxi, et al. "A multiscale edge detection algorithm based on wavelet domain vector hidden Markov tree model." *Pattern Recognition* 37.7 (2004): 1315-1324.
[6]. Nowak, Robert D. "Multiscale hidden Markov models for Bayesian image analysis." Bayesian Inference in Wavelet-Based Models. Springer, New York, NY, 1999. 243-265..
[7]. Fan, Guoliang, and Xiang-Gen Xia. "Improved hidden Markov models in the wavelet-domain." *IEEE Transactions on Signal Processing* 49.1 (2001): 115-120.

[8]. Crouse, Matthew S., Robert D. Nowak, and Richard G. Baraniuk. "Wavelet-based statistical signal processing using hidden Markov models." IEEE Transactions on signal processing 46.4 (1998): 886-902.
[9]. Rabiner, Lawrence R. "A tutorial on hidden Markov models and selected applications in speech recognition." *Proceedings of the IEEE* 77.2 (1989): 257-286.
[10].     Choi, Hyeokho, and Richard G. Baraniuk. "Multiscale image segmentation using wavelet-domain hidden Markov models." IEEE Transactions on Image Processing 10.9 (2001): 1309-1321.